\documentclass[a4paper,fleqn]{cas-dc}

\usepackage[authoryear,longnamesfirst]{natbib}
\usepackage[figuresright]{rotating}
\usepackage{color}

\definecolor{darkmagenta}{rgb}{0.55, 0.0, 0.55}

\definecolor{darkgreen}{rgb}{0,0.5,0}

\usepackage{ulem} 

\usepackage{tabularx}
\usepackage{adjustbox}
\usepackage{graphicx}
\usepackage{subcaption}
\usepackage{array}

\def\tsc#1{\csdef{#1}{\textsc{\lowercase{#1}}\xspace}}
\tsc{WGM}
\tsc{QE}

\begin{document}
\let\WriteBookmarks\relax
\def\floatpagepagefraction{1}
\def\textpagefraction{.001}

\shorttitle{Transitioning a Project-Based Course between Onsite and Online}    

\shortauthors{D. Suciu, S. Motogna, A. Molnar}  

\title [mode = title]{Transitioning a Project-Based Course between Onsite and Online. An Experience Report}



%

\author{Dan Mircea Suciu}[orcid=0000-0002-5958-419X]

\cormark[1] 

\ead{dan.suciu@ubbcluj.ro}

\ead[url]{http://www.cs.ubbcluj.ro/~dsuciu}

\credit{Conceptualization, Methodology, Data collection, Qualitative and quantitative analysis; Writing - original draft, review and editing}

\affiliation[1]{organization={Babes-Bolyai University, Department of Computer Science},
          addressline={M. Kogalniceanu 1}, 
            city={Cluj-Napoca},
          postcode={400084}, 
           country={Romania}}

\author{Simona Motogna}[orcid=0000-0002-8208-6949]

\cormark[1]


\ead{simona.motogna@ubbcluj.ro}

\ead[url]{http://www.cs.ubbcluj.ro/~motogna}

\credit{Methodology, Data collection and qualitative analysis; Writing - original draft, review and editing}


\author{Arthur-Jozsef Molnar}[orcid=0000-0002-4113-2953]

\cormark[1]


\ead{arthur.molnar@ubbcluj.ro}

\ead[url]{http://www.cs.ubbcluj.ro/~arthur}

\credit{Data collection and qualitative analysis; Writing of the original draft, Threads to validity, review and editing}



\cortext[1]{Corresponding author}



\begin{abstract}
We present an investigation regarding the challenges faced by student teams across four consecutive iterations of a team-focused, project-based course in software engineering. The studied period includes the switch to fully online activities in the spring of 2020, and covers the return to face-to-face teaching two years later. We cover the feedback provided by over 1,500 students, collected in a free-text form on the basis of a survey. A qualitative research method was utilized to discern and examine the challenges and perceived benefits of a course that was conducted entirely online. We show that technical challenges remain a constant in project-based courses, with time management being the most affected by the move to online. Students reported that the effective use of collaborative tools eased team organization and communication while online. We conclude by providing a number of action points regarding the integration of online activities in face-to-face course unfolding related to project management, communication tools, the importance of teamwork, and of active mentor participation. 
\end{abstract}


\begin{highlights}
\item Impact analysis of changes from in-person to online and back to in-person education
\item Research of project-based course challenges pre-, during and post-pandemic.
\item Online-to-onsite transfer affects mainly student teams organization

\end{highlights}


\maketitle

\section{Introduction}
\label{SIntro}

The software industry and society pose high expectations on the professional skills of university graduates. Those skills include technical abilities to develop and maintain complex software systems but also soft skills, such as teamwork. Universities need to answer these demands while considering the dynamic changes of the domain. The pandemic put extra pressure on the education system, as fast adaptation was required. Now, as things come back to normal, there two important consequences that must be considered. On one hand, industry working practices have changed, as many companies adopted remote or hybrid work. On the other hand, universities must learn from the experience of fully online education.

Project-based courses in which students are required to work in teams and walk through all development stages have already gained the reputation of developing essential skills for future professionals \citep{FAH13,BRS17}. This type of courses present additional issues from an educational perspective, as they combine acquiring of technical and soft skills by students with introducing the evaluation of individual students in a group or team setup. 

\cite{Gary17} discussed the mismatch between the actual challenges and those expected by the faculty staff related to the interaction between students and the teaching staff, or between the students themselves. Similarly, \cite{IF19} described the existence of a significant gap between student expectations on teamwork and the experience in retrospect, proposing setting realistic expectations at project onset, focusing on team organization and informal meetings, and emphasizing the role of leadership. Several studies \citep{Souza,FCS,HAC18,BRS17} contributed to the topic, with most of them addressing student well being or discussing challenges associated with particular courses. Our perception is that a more exhaustive investigation of these challenges and a classification of them can lead to better solutions, thus contributing to more effective project based learning.

We propose an exploratory study in which data collected from several years of a project-based course reports about the challenges faced by students. The data reflects the changes from the face to face setting to online and then back to face to face activities, thus being a valuable indicator for changes associated with transitioning between online and onsite education. 

We looked at challenges as defined by the Cambridge Dictionary\footnote{\url{https://dictionary.cambridge.org/dictionary/english/challenge}}, as \textit{"the situation of being faced with something that needs great mental or physical effort in order to be done successfully and therefore tests a person's ability"}

The main goal of the study is to compare challenges students have faced during a period of four years and to gain insight on the appropriateness of perpetuating practices installed in online activities to the face to face environment. The investigation is based on data collected through surveys and processed through thematic analysis.

The key findings of our approach can be succinctly summarized as follows:

\begin{itemize}
\item Comparative investigation of results obtained through thematic analysis
\item Insights regarding student perceptions on challenges in team projects, both online and onsite
\item Action points to be considered for project-based courses in which online activities can be successfully integrated in an onsite setting.
\end{itemize}


The paper is organized as follows: the next section presents related work. Section \ref{SOverview} provides the context in which our study was executed and the used methodologies. Results are discussed in Section \ref{Results}, followed by the action points derived from our findings. Threats to validity are discussed in Section \ref{SThreats}, and the final section is dedicated to conclusions and an overview of the planned continuation of the study.

\section{Related work}
\label{SSRelatedwork}

The move to online teaching in the spring of 2020 was sudden, with hindsight revealing in many cases a lack of preparedness from both the teaching staff and students. The consequences of this change were extensively reported in the scientific literature. We believe it is worthwhile to also examine the effect that the return to traditional, on-site activities had on teaching opportunities and the challenges faced by students.

A large number of contributions can be found in the literature related to switching academic courses to online due to the Covid-19 pandemic. These reported experiences from both the side of the academic staff (such as \cite{Moto20}) as well as from the students' perspective, from different areas of the world: the USA (\cite{Means}), Asia (\cite{Hazeyama2022,Yamada20}), the Middle East (\cite{Hassan21,Fitoussi21}), Latin America (\cite{Salas22}), Australia (\cite{Kanij20}) and Europe (\cite{Barr20}). All these contributions focused on the transition to online activities and how to adapt an educational model that for decades functioned strictly on site, and mostly concentrated on the students' experiences with regards to challenges, difficulties, perceptions or reflections.

Students' satisfaction, experiences and technology access were the focus in \cite{Means}, a large study addressed to over 1,000 students. The results were formulated as challenges to transitioning to online education, analyzed based on their severity, and mainly related to the students' personal motivation.

\cite{Hazeyama2022} reported that project based learning and the use of collaborative tools such as GitHub have been successful in overcoming difficulties specific to online education. A similar experiential study \citep{Yamada20} revealed that developing appropriate online educational tools can assure a good student - professor interaction and ease the evaluation process.

A comprehensive statistical analysis of students' perceptions and satisfaction is presented in \cite{Hassan21}. The 328 respondents filled in an anonymous questionnaire reporting on their perception regarding personal experiences, availability of services and facilities as the main determinants of their satisfaction. The outcome of this study stated that perceptions and satisfaction have been the main contributors to student motivation for academic performance.

\cite{Salas22} presented a systematic literature review of approaches taken in Latin America for transitioning to online education, investigating challenges and opportunities. Their initial search yielded 717 research papers that were filtered according to inclusion and exclusion criteria, with duplicates eliminated. Out of the 18 remaining papers, only five reported data for a student sample size larger than 100, with only one study including over 1,000 students. Reported results identified stress and depression as the main challenges facing students, with increased workload being the main challenge of the teaching staff. 

\cite{Kanij20} reported the authors' experiences regarding a Software Engineering course and summarized the identified challenges from an educator's perspective: stressful transition, student engagement, asynchronous learning, offline demos and assessment.

Since team organization often challenges students working in a collaborative environment, we examined the literature focused on team organization and process. \cite{RAF18} reported on efforts to stimulate teamwork and communication in team projects at the University of Milano - Bicocca. Teams employed GitHub as a collaborative platform and used SonarQube to assess the quality of developed software. From a technical skills perspective, students appreciated using tools that are widely employed in the industry, a finding that we mirrored in both our previous \citep{enase21} and current study. In \citep{Heberle2018}, authors reported on eight iterations of a project-driven course, covering the implementation of over 100 projects, each estimated at 1,000 person-hours. Similar to our experience, the course employed both teaching staff as well as outside collaborators filling the role of customers, in order to provide a more realistic experience. This was shown to help students acknowledge their weaknesses, improve communication, and help them integrate into future teams. Authors of \citep{Masood2018} examined student challenges within two consecutive iterations of a course based on the application of Agile practices in a project-based course, with findings that we subsequently validated in our own work \citep{springer2022}. The organizational and team-related issues occurring in the development of an industrial application as part of a master's level course were detailed in \citep{rai22b}. Authors reported on their experience and proposed activities to be carried out before, during and after the course in order to drive student interest and engagement.  

In \citep{csedu22}, authors focused their investigation on team organization and collaboration in an online software engineering course. They discovered that although most teams supported working in a distributed environment, they preferred an initial onsite phase in order to set up the project. The lessons learned include the use of common terminology and tools, as well as carrying out frequent reviews of application design.

Student course satisfaction before, during and after the switch to online activities caused by the pandemic was reported in \citep{cor22}. Authors used the institution provided 9-item questionnaire to analyze 7 course iterations, two of which were before the pandemic and one after. They observed course satisfaction to gradually decline during the pandemic, with levels returning once in-person activities were resumed. One key observation regarded the declining number of students and provided feedback, which was explained by lower student engagement due to \textit{"zoom fatigue"} and the monotonous nature of strictly online activities. 

Compared to the reported studies, our approach is focused on the software engineering domain and reports student feedback gathered over four consecutive course iterations. The large number of responses, representing the experience of over 1,500 students, as well as their non-anonymous nature help differentiate our study from the existing literature. We detail the learning objectives and outcomes of the targeted course, and consider the related challenges in both the onsite and online environments. The impact of Covid-19, the transition to online activities and then the recommencement of face to face activities are studied as an important factor that influence student perceived challenges as well as the results of their work. The present paper represents the continuation of our work regarding the impact and effect of the changes that were forced by the Covid-19 pandemic on the higher education system in general and the students in particular. We started by examining student perception of software engineering team projects \cite{enase21}, as well as the adoption of Agile practices \cite{springer2022}; these initial works employed student feedback from the 2019-2020 iteration of the course that was fully on site and unaffected by the subsequent pandemic. The adaptation to fully online teaching and evaluation was examined in \cite{Moto20}, where quantitative and qualitative approaches were combined in order to examine teaching staff adaptation to the new conditions. The students' perspective regarding the challenges of moving project-based courses online were described in \cite{mot22}. In our present report, we complete the circle by examining the student reported challenges during the transition back to onsite teaching. In order to keep results comparable across the four iterations of the course, we kept the same structure and data collection methodology for the student survey. Our current research is based on responses covering a sample size of more than 1,500 students and covers one course iteration before any pandemic-related changes, two iterations affected by the pandemic, with the final iteration taking place after all Covid-19 restrictions have been lifted. We focused our research on a more detailed analysis of student reported challenges, as our previous works \cite{enase21,springer2022} have illustrated them to be the area of most interest.


\section{Overview and design of the study}
\label{SOverview}

This section outlines the study design (research objectives with research questions), information about data collection and data analysis.

Our study relies on the Team Project course offered to third-year Computer Science students in the Mathematics and Computer Science Faculty of our University. Course details are presented in Section \ref{StudyBackground}

\subsection{Purpose of the study}
\label{SDesignPurpose} 

We organized our research and carried out a qualitative analysis of how students perceived the difficulties and benefits of working on collaborative projects both online and in-person. We quantified our primary objective by framing it in terms of the two perspectives listed below: 
\newline
\newline
\textbf{$RQ_{1}$}. \textit{How does transitioning between online and face-to-face learning impact the challenges faced by students in project-based courses?}

We investigated the challenges that were associated with both hard (organizational and technical) and soft skills, based on how  students actually interpreted those challenges. In order to provide an appropriate response to this question, we took steps to ascertain the existence of any widespread challenges and to characterize their frequency, degree of difficulty, and degree of variance during the shifts from onsite to online and then from online to onsite educational contexts.
\newline
\newline
\textbf{$RQ_{2}$}.\textit{Which are the effects of online learning experience on the perception of face-to-face learning?}

We explored the data that was collected regarding opinions about online activities in order to gain a better understanding of the benefits and drawbacks of the online approach from the students' point of view, as well as to determine whether the shift back to an onsite method of working has altered this perspective. 

\subsection{Study background}
\label{StudyBackground} 

The \textit{Team Project} course is offered in the 3$^{rd}$ year of undergraduate studies, when students have already gained experience with programming languages, fundamental software development and IDEs. The main goal of the course is to prepare software engineering students for industry jobs by helping them learn and combine technical and soft skills when working in a team. As general assignment, students are asked to develop a software application of their choice during a 14-week period in teams supervised by mentors from the software industry.

The course was designed with three main principles to be integrated: (i) \textit{Project based:} the software application is the kernel of the course and the main asset in the evaluation of students; (ii) Student teams are \textit{mentored} by experienced professional in order to simulate the experience of a real life project; (iii) Students should also gain soft skills through organizing \textit{workshops} that were especially designed for them. Concurrently, students participated in four workshops on communication, public speaking, the Agile mindset, and entrepreneurship\footnote{
course syllabus: \url{https://www.cs.ubbcluj.ro/files/curricula/2021/syllabus/IE\_sem5\_MLR5012\_ro\_dsuciu\_2021\_6331.pdf}}.

At the end of the semester, each team made a 15 minute live presentation, including a demo of the application in front of the teaching staff and the mentors. 

The data used in our study corresponds to four consecutive academic years, between 2019 and 2023. Data on the number of students, mentors and teams involved each year are presented in Table \ref{table:course_attendance}, and totaled 1,536 students grouped into 159 teams. The table also specifies whether the courses were conducted in online or face-to-face format.

\begin{table}
  \caption{Team Project course attendance}
  \label{table:course_attendance}
  \begin{tabular}{ccccc}
    \toprule
    Year&Students&Teams&Mentors&Type\\
    \midrule
    2019-2020 & 492 & 49 & 15 & face-to-face \\
    2020-2021 & 317 & 35 & 12 & online \\
    2021-2022 & 410 & 39 & 15 & online \\
    2022-2023 & 317 & 36 & 15 & face-to-face \\
  \bottomrule
\end{tabular}
\end{table}

\begin{table*}[t]
  \caption{Summary of categories and topics for student challenges in all editions of the course, determined through open coding }
  \label{tab:all_topics} \centering
  \begin{tabular}{|p{2.5cm}|p{4.8cm}|p{3cm}|p{5.2cm}|} \hline \cellcolor{lightgray!25}\textbf{Categories}&\cellcolor{lightgray!25}\textbf{Organizational}&\cellcolor{lightgray!25}\textbf{Technical}&\cellcolor{lightgray!25}\textbf{Soft skills}\\ \hline
    \multirow{3}{*}{\textbf{Topics 2019-2020}}&\cellcolor{pink!25}Time management, task management, teamwork, collaboration, technology choice&\cellcolor{Yellow!25}Technical skills, over-engineering&\cellcolor{LimeGreen!25}Involvement, communication, teamwork \\ \hline
    \multirow{3}{*}{\textbf{Topics 2020-2021}}&\cellcolor{pink!25}Team organization, time management, teamwork, team synchronization, project organization&\cellcolor{Yellow!25}New technology, over-engineering, technical issues &\cellcolor{LimeGreen!25}Lack of face-to-face communication, divergent vision, leadership, lack of communication, lack of engagement\\ \hline
    \multirow{4}{*}{\textbf{Topics 2021-2022}}&\cellcolor{pink!25}Ineffective teamwork/team organization, defective time management, lack of vision, inappropriate project organization &\cellcolor{Yellow!25}Lack of technical skills&\cellcolor{LimeGreen!25}Lack of effective communication, lack of leadership, lack of engagement\\ \hline
    \multirow{3}{*}{\textbf{Topics 2022-2023}}&\cellcolor{pink!25}Ineffective teamwork/team organization, defective time management, inappropriate project organization &\cellcolor{Yellow!25}Lack of technical skills &\cellcolor{LimeGreen!25}Lack of effective communication, lack of engagement, lack of leadership \\ \hline
\end{tabular}
\end{table*}

Between iterations of the considered course, organizational and structural modifications occurred. In its first iteration (2019-2020), the course was required for all Computer Science and Mathematics and Computer Science students in the Faculty.
Later in 2020, this course became an elective one, providing students the option of selecting this program or selecting an individual research topic to pursue.

In addition, beginning with the second iteration of the course, we implemented the Code of Talent \citep{cot2022} microlearning platform, which enabled student teams to share project topics, intermediate demo videos, project documentation, and retrospective outcomes. Moreover, the platform offered gamification through the assignment of learning points and knowledge badges, which improved their engagement throughout the semester. In this approach, we established an environment that fostered the sharing of ideas, enabled contact across teams, and boosted the motivation and competitiveness of team members.

Due to the pandemic-imposed restrictions that were enforced throughout the second and third iterations of the course, all team meetings, mentoring sessions and seminars were conducted online. 


\subsection{Data collection}
\label{SDataCollection} 

The data collection process was conducted as part of the Team Project's suggested learning activities. Given the number of participants in the course, we opted for a survey since it was straightforward to organize and time-efficient. The type of survey we chose is exploratory, in the form of open ended questions. As part of the process, students were given some recommendations on how to respond: first of all, the answers should contain details and descriptions such that each answer was consistent in terms of size and content; secondly, each team should submit one response, which means that students reflected about their work and then formulated the answer as a collective opinion.

Each team was asked to reflect on their experiences and then invited to reply to the following survey (one response per team) following the conclusion of technical project work, under the guidance of the mentors. The questions were correlated with the course outcomes:
\begin{itemize}
\item \textbf{Process}. \textit{Briefly describe the development process or the methodology implemented by the team.}
\item \textbf{Time management}. \textit{Characterise how the initial planning was adhered to during the course of the project.}
\item \textbf{Challenges}. \textit{What were the 3 most important challenges encountered and how were they addressed (if applicable)?}
\item \textbf{Lessons learned}. \textit{What are the most relevant lessons learned during the development of the software solution?}
\item \textbf{Online}. \textit{Name three aspects that made it easier to organize the team and manage the project online.} (available only for 2021-2022 academic year)
\item \textbf{Online}. \textit{Do you consider that participating online in meetings with mentors or workshops would have been more appropriate or, on the contrary, would have involved more disadvantages?} (available only for 2022-2023 academic year)
\end{itemize}
All responses were free-text, and students were encouraged to write thorough descriptions and to reflect on meaningful experiences. This paper focuses on the responses to the \textbf{Challenges} and \textbf{Online} parts since they provided the most pertinent information regarding the pros and cons of co-located versus distributed/virtual work. We created and published an open-data package that contains the student feedback to these sections for all iterations of the course \citep{jss2023}.

In terms of collected data, not all teams responded to the \textbf{Challenges} section of the questionnaire. Three and two teams skipped this section during the 2019-2020 and 2021-2022 academic years, respectively. Similarly, for the 2022-2023 interval, we received responses from only 26 teams, with 25 of which provided answers to the question about online appropriateness. 

\subsection{Methodology}
\label{SMethod}

\begin{table}[h!]
    \caption{Number of occurrences of topics for challenges identified through open coding in 2019-2021}
    \label{fig:challenges-categories1}
    \begin{tabular}{|p{4cm}|c|c|}
    \hline
    \cellcolor{lightgray!25}\textbf{Topics 2019 - 2020} &\cellcolor{lightgray!25}\textbf{Count}&\cellcolor{lightgray!25} \textbf{Category}\\ \hline
    \cellcolor{pink!25}Time management&\cellcolor{pink!25}16&\cellcolor{pink!25}Organizational\\
    \cellcolor{pink!25}Task management&\cellcolor{pink!25}14&\cellcolor{pink!25}Organizational\\
    \cellcolor{pink!25}Lack of technology stack agreement&\cellcolor{pink!25}\multirow{2}{*}{8}&\cellcolor{pink!25} \multirow{2}{*}{Organizational}\\
    \cellcolor{pink!25}Teamwork&\cellcolor{pink!25}4&\cellcolor{pink!25}Organizational\\
    \cellcolor{pink!25}Time synchronization&\cellcolor{pink!25}4&\cellcolor{pink!25}Organizational\\
    \cellcolor{pink!25}Management of expectations&\cellcolor{pink!25}3&\cellcolor{pink!25}Organizational\\
    \cellcolor{pink!25}Co-location&\cellcolor{pink!25}2&\cellcolor{pink!25}Organizational\\
    \cellcolor{pink!25}Define project idea&\cellcolor{pink!25}2&\cellcolor{pink!25}Organizational\\
    \cellcolor{pink!25}Effective collaborations&\cellcolor{pink!25}4&\cellcolor{pink!25}Organizational\\
    \cellcolor{pink!25}Team organization&\cellcolor{pink!25}2&\cellcolor{pink!25}Organizational\\
    \cellcolor{pink!25}Team coordination&\cellcolor{pink!25}2&\cellcolor{pink!25}Organizational\\ 
    \cellcolor{pink!25}Communication with the mentor &\cellcolor{pink!25}\multirow{2}{*}{1}&\cellcolor{pink!25}\multirow{2}{*}{Organizational}\\
    \cellcolor{pink!25}Confusion regarding team roles&\cellcolor{pink!25}\multirow{2}{*}{1}&\cellcolor{pink!25}\multirow{2}{*}{Organizational}\\ \hline
    \cellcolor{LimeGreen!25}Lack of involvement&\cellcolor{LimeGreen!25}10&\cellcolor{LimeGreen!25}Soft skills\\
    \cellcolor{LimeGreen!25}Effective communication&\cellcolor{LimeGreen!25}12&\cellcolor{LimeGreen!25}Soft skills\\
    \cellcolor{LimeGreen!25}Pressure to work in a team&\cellcolor{LimeGreen!25}1&\cellcolor{LimeGreen!25}Soft skills\\ \hline
    \cellcolor{Yellow!25}Lack of technical skills&\cellcolor{Yellow!25}25&\cellcolor{Yellow!25}Technical\\
    \cellcolor{Yellow!25}Over-engineering&\cellcolor{Yellow!25}1&\cellcolor{Yellow!25}Technical\\ \hline
    \cellcolor{lightgray!25}\textbf{Topics 2020 - 2021} &\cellcolor{lightgray!25}\textbf{Count}&\cellcolor{lightgray!25} \textbf{Category}\\ \hline
    \cellcolor{pink!25}Time management&\cellcolor{pink!25}8&\cellcolor{pink!25}Organizational\\
    \cellcolor{pink!25}Project organization&\cellcolor{pink!25}13&\cellcolor{pink!25}Organizational\\
    \cellcolor{pink!25}Team organization&\cellcolor{pink!25}16&\cellcolor{pink!25}Organizational\\
    \cellcolor{pink!25}Team sync&\cellcolor{pink!25}15&\cellcolor{pink!25}Organizational\\
    \cellcolor{pink!25}Teamwork&\cellcolor{pink!25}2&\cellcolor{pink!25}Organizational\\ \hline
    \cellcolor{LimeGreen!25}Divergent vision&\cellcolor{LimeGreen!25}3&\cellcolor{LimeGreen!25}Soft skills\\
    \cellcolor{LimeGreen!25}Lack of communication&\cellcolor{LimeGreen!25}5&\cellcolor{LimeGreen!25}Soft skills\\
    \cellcolor{LimeGreen!25}Lack of engagement&\cellcolor{LimeGreen!25}3&\cellcolor{LimeGreen!25}Soft skills\\
    \cellcolor{LimeGreen!25}Lack of face-to-face communication&\cellcolor{LimeGreen!25}\multirow{2}{*}{5}&\cellcolor{LimeGreen!25}\multirow{2}{*}{Soft skills}\\
    \cellcolor{LimeGreen!25}Leadership&\cellcolor{LimeGreen!25}1&\cellcolor{LimeGreen!25}Soft skills\\ \hline
    \cellcolor{Yellow!25}New technologies&\cellcolor{Yellow!25}20&\cellcolor{Yellow!25}Technical\\
    \cellcolor{Yellow!25}Over engineering&\cellcolor{Yellow!25}1&\cellcolor{Yellow!25}Technical\\
    \cellcolor{Yellow!25}Technical issues&\cellcolor{Yellow!25}2&\cellcolor{Yellow!25}Technical\\ \hline
    \end{tabular}
\end{table}

The selected research methodology was \textit{qualitative analysis}, specifically in the form of thematic analysis, as suggested by \citep{Braun2019} for studies involving surveys with open-ended questions and responses. Thematic analysis was successfully used in tackling software engineering problems \citep{Cruzes,Peggy}, and our methodology followed the recommendations from  \citep{Kiger,ACM}. 

Reflexive thematic analysis was applied \citep{Kiger} since we considered it best suited for our case, based on the following observations: themes are conceptualized as meaning based patterns, the coding process is open, flexible and iterative and the goal is to provide a systematic interpretation of data supported by arguments deduced from the data.

The open coding followed by thematic analysis was an iterative process in which all three authors played different roles in turn (coding, conceptualization, categorization and verification) in order to assure the validity of the results. The steps performed can be summarized in:
\begin{itemize}
\item Initial step of coding in which keywords were detected in the free text, then topics were associated to them;
\item The second step consisted of conceptualization and categorization of the detected topics;
\item Each of the first two steps was performed by two coders, and during the third step the third coder checked the topics, created categories and then proposed a merging process, which was followed by debate and mutual agreement between all three researchers involved in the process.
\end{itemize}

As an example, starting from the text “A huge challenge was synchronizing our schedules so that we could work together” we identified “synchronizing schedule” as a keyword, which we tagged as topic “time synchronization”, and then in the second step, considering also the keywords and topics detected from all answers the category “organizational” was created, and in the third step the topic was merged to “time management”.

Table \ref{tab:all_topics} summarizes the process: the columns correspond to the merged categories, and the topics associated to each category are represented by year. An important remark regarding the process is that the three steps described above were applied independently for each data set, respectively for each iteration of the course in an academic year, and only afterwards were the categories verified against each other.


The three categories of challenges, namely organizational, technical and related to soft skills, are in agreement with those from similar studies: \cite{IF19} referred to team, skills, process and environment, \cite{rai18} identified challenges as collaboration, teamwork and tools, while recent studies, such as \cite{Wloda} and \cite{Presler} mentioned communication, team management, teamwork and tasks.

\subsection{Data analysis}
\label{SDataAnalysis} 

For the academic years 2019-2020 and 2020-2021 we obtained the results shown in Table \ref{fig:challenges-categories1}. To obtain these results we applied the Open Coding technique, various aggregations of which were presented in two previous papers \citep{enase21,mot22}) and briefly described in section \ref{SMethod}. In order to continue the analysis of the data collected in subsequent years, a mapping of the challenge topics was necessary. Thus, we compared the topics we found in both years and put them in correspondence, obtaining in a first step the situation shown in Table \ref{fig:challenges-mapping}.

\begin{table}
    \caption{Challenge topics mapping}
    \label{fig:challenges-mapping}
    \begin{tabular}{|p{3.9cm}|p{3.9cm}|}
    \hline
    \multicolumn{2}{|c|}{\cellcolor{lightgray!25}\textbf{Organizational challenges}} \\ \hline
    \cellcolor{lightgray!25}\textbf{2019 - 2020}&\cellcolor{lightgray!25}\textbf{2020 - 2021}\\ \hline
    \cellcolor{pink!25}Task management&\cellcolor{pink!25}Project Organization\\ \hline
    \cellcolor{pink!25}Effective collaboration, teamwork&\cellcolor{pink!25}\multirow{2}{*}{Teamwork}\\ \hline
    \cellcolor{pink!25}Communication with the mentor&\cellcolor{pink!25}\\ \hline
    \cellcolor{pink!25}Time synchronization, time management&\cellcolor{pink!25}\multirow{2}{*}{Time management}\\ \hline
    \cellcolor{pink!25}Team Organization, co-location, confusion regarding team roles, team coordination&\cellcolor{pink!25}\multirow{4}{*}{Team Organization}\\ \hline
    \cellcolor{pink!25}&\cellcolor{pink!25}Team sync\\ \hline
    \cellcolor{pink!25}Define project idea, lack of technology stack agreement, management of expectations&\cellcolor{pink!25}\\ \hline
    \multicolumn{2}{|c|}{\cellcolor{lightgray!25}\textbf{Soft Skills challenges}}\\ \hline
    \cellcolor{lightgray!25}\textbf{2019 - 2020}&\cellcolor{lightgray!25}\textbf{2020 - 2021}\\ \hline
    \cellcolor{LimeGreen!25}Pressure to work in a team&\cellcolor{LimeGreen!25}\\ \hline
    \cellcolor{LimeGreen!25}&\cellcolor{LimeGreen!25}Divergent vision\\ \hline
    \cellcolor{LimeGreen!25}\multirow{3}{*}{Effective communication}&\cellcolor{LimeGreen!25}Communication, lack of face-to-face communication, lack of communication\\ \hline
    \cellcolor{LimeGreen!25}Lack of involvement&\cellcolor{LimeGreen!25}Lack of engagement\\ \hline
    \cellcolor{LimeGreen!25}&\cellcolor{LimeGreen!25}Leadership\\ \hline
    \multicolumn{2}{|c|}{\cellcolor{lightgray!25}\textbf{Technical challenges}}\\ \hline
    \cellcolor{lightgray!25}\textbf{2019 - 2020}&\cellcolor{lightgray!25}\textbf{2020 - 2021}\\ \hline
    \cellcolor{Yellow!25}Lack of technical skills&\cellcolor{Yellow!25}New technologies, technical issues\\ \hline
    \cellcolor{Yellow!25}Over-engineering&\cellcolor{Yellow!25}Over engineering\\ \hline
    \end{tabular}
\end{table}

We considered the following changes to the initially identified challenges:
\begin{itemize}
\item \textit{Communication with the mentor} is more a part of leadership skills. We have renamed this topic to \textit{Lack of leadership} to more accurately describe a challenge.
\item \textit{Effective collaboration} from \textit{Organizational} challenges and \textit{Pressure to work in a team} from \textit{Soft Skills} challenges are topics of \textit{Ineffective teamwork}, and we considered them as being covered by this particular topic.
\item In \textit{Organizational} challenges there are 3 topics that are closely related to each other - (\textit{Management of expectations}, \textit{Lack of technology stack agreement} and \textit{Define project idea}) and could be grouped under the name \textit{Lack of vision}. We have also included here \textit{Divergent vision}, that appears as part of the \textit{Soft Skills} challenges.
\item We also considered \textit{Team sync} as part of the topic \textit{Defective team organization}.
\item In the end, we considered that in an Agile team there is a lot of confusion between teamwork and team organization, since we expect to build a self-organizing and self-coordinating team. Therefore, we merged these two topics in one called \textit{Ineffective teamwork/team organizaton}.
\end{itemize}

After restructuring, the final challenge topics are shown in Table \ref{fig:challenges_final_categories}. We used these categories and topics for performing thematic analysis on data collected in the 2021-2022 and 2022-2023 academic years. The results are presented in Table \ref{fig:challenges-statistics2withoutChatGPT}. 

\begin{table}[h]
    \caption{Final list of challenge categories and topics}
    \label{fig:challenges_final_categories}
    \begin{tabular}{|l|}
    \hline
    \cellcolor{lightgray!25}\textbf{Organizational Challenges}\\ \hline
    \cellcolor{pink!25}Inappropriate project organization\\
    \cellcolor{pink!25}Defective time management\\
    \cellcolor{pink!25}Ineffective teamwork/Team organization\\
    \cellcolor{pink!25}Lack of vision\\ \hline
    \cellcolor{lightgray!25}\textbf{Soft Skills Challenges}\\ \hline
    \cellcolor{LimeGreen!25}Lack of leadership\\
    \cellcolor{LimeGreen!25}Lack of effective communication\\
    \cellcolor{LimeGreen!25}Lack of engagement\\ \hline
    \cellcolor{lightgray!25}\textbf{Technical Challenges}\\ \hline
    \cellcolor{Yellow!25}Lack of technical skills\\
    \cellcolor{Yellow!25}Over engineering\\ \hline
    \end{tabular}
\end{table}

\begin{table*}[h]
    \caption{Number of topics per category/academic year/analysis source after topics restructuring}
    \label{fig:challenges-statistics2withoutChatGPT}
    \begin{tabular}{|p{4cm}|c|c|c|c|} \hline
    \cellcolor{lightgray!25}\textbf{Academic year}&\cellcolor{lightgray!25}\textbf{Organizational (4)}&\cellcolor{lightgray!25}\textbf{Soft Skills (3)}&\cellcolor{lightgray!25}\textbf{Technical (2)}&\cellcolor{lightgray!25}\textbf{Source}\\  \hline
    \textbf{2019 - 2020}&\cellcolor{pink!25}63&\cellcolor{LimeGreen!25}23&\cellcolor{Yellow!25}26&\multirow{1}{*}{46 teams, 4222 words}\\
    \hline
    \textbf{2020 - 2021}&\cellcolor{pink!25}57&\cellcolor{LimeGreen!25}14&\cellcolor{Yellow!25}23&\multirow{1}{*}{35 teams, 4398 words}\\
    \hline
    \textbf{2021 - 2022}&\cellcolor{pink!25}37&\cellcolor{LimeGreen!25}28&\cellcolor{Yellow!25}26&\multirow{1}{*}{37 teams, 4765 words}\\
    \hline
    \textbf{2022 - 2023}&\cellcolor{pink!25}20&\cellcolor{LimeGreen!25}10&\cellcolor{Yellow!25}11&\multirow{1}{*}{25 teams, 1963 words}\\
    \hline
    \end{tabular}
\end{table*}

\section{Results and discussion}
\label{Results}

This section contains the results and insights regarding our study perspectives. 
\newline
\newline
\textbf{$RQ_{1}$}. \textit{How does transitioning between online and face-to-face learning impact the challenges faced by students in project-based courses?}

A project-based course is more likely than other courses to require students to step outside their comfort zone. The reason for this is that a student's performance is no longer solely dependent on his or her own efforts, but also on how his or her teammates contribution towards achieving the goals.

Similarly, at the individual level, students will test not only their technical skills but also their soft skills (communication, collaboration, etc.) and organizational skills (following processes and methodologies and using tools to coordinate activities).

Therefore, it was not entirely unexpected that the analysis of the challenges faced by the students led to the identification of three categories: organizational, soft skills, and technical. 

Although we did undertake an examination of all of the challenges that were reported by students, the objective of this study was to determine the variations that were influenced by the online or onsite approach to the course.

Figure \ref{fig:challenge_4_years} represents the results of applying open coding with thematic analysis on the challenges of the four academic years. These results led us to the following observations:

\begin{figure*}[h]
    	\centerline{
        \includegraphics[scale=0.3]
        {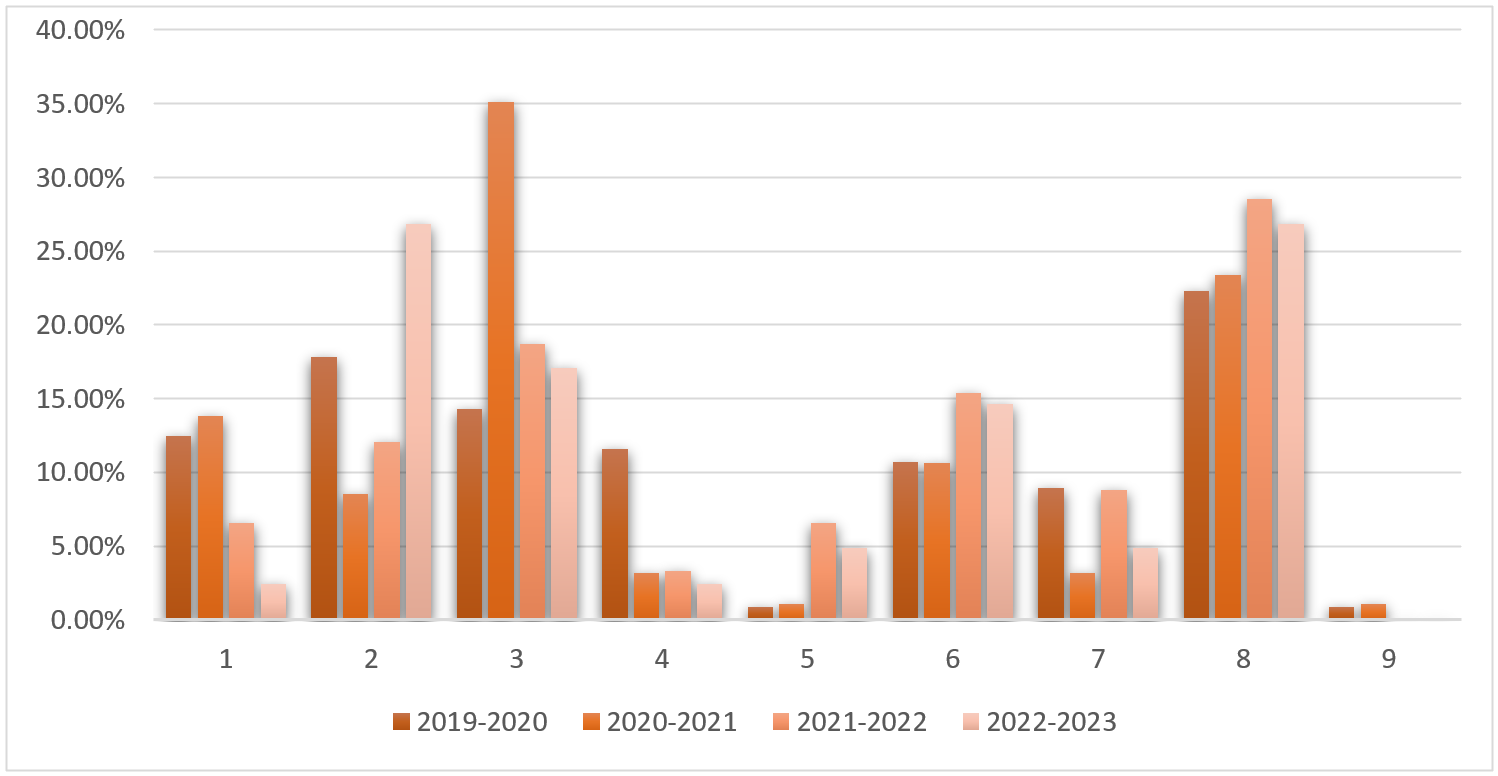}  
    }
\caption{Distribution of challenge topics for all four academic years based on thematic analysis (1 - Inappropriate project organization, 2 - Defective time management, 3 - Defective team organization, 4 - Lack of vision, 5 - Lack of leadership, 6 - Lack of effective communication, 7 - Lack of engagement, 8 - Lack of technical skills, 9 - Over engineering)}
    \label{fig:challenge_4_years}
\end{figure*}

\textbf{Technical challenges are a constant}.
Significant proportions of the difficulties reported by students (between 22.32\% and 28.57\% ) were associated with the knowledge, understanding, and application of the technologies chosen for the implementation of software solutions. This was attributable in part to variances in the individual knowledge of team members (\textit{''...most of my colleagues on the front were not used to working on React...'', ''...not all of us were proficient in the chosen tech stack...''}), but also to the fact that some teams elected to utilize new technologies and familiarized themselves with them throughout the development of the project (\textit{''...Using technologies we were not familiar with before...'', ''...with quite new technologies, we had some technical challenges'',''...The enthusiasm at the beginning of the project pushed us to a rather ambitious stack...''}).

Our data indicates that the shift from on-site to online and vice versa had no major effect on this aspect.

\textbf{Time management and project organization capacities were most affected}.
The time management difficulties revealed the most important differences.
On the one hand, we observed a certain percentage rise in these obstacles with the transition from on-site to online delivery.
There were a number of reasons for this, including familiarity with online communication tools and the fact that the switch from face-to-face to online corresponded with an optional course structure change (there was a sudden increase of project organization difficulties, from 14.29\% in onsite context to 35.11\% in online).
The latter resulted in teams of students with varying learning schedules who had a more difficult time finding a common working period (\textit{''...some of us worked at night, others in the morning, each of us had their own personal schedule...'', ''...Find a common time for synchronization meetings...''}).
Delays resulting from the inability to synchronize activities led to scheduling issues (\textit{''...A first challenge was to synchronise the team members...'', ''...not all of us managed to get into the meetings as planned...''}).  



%
At the same time, it turns out, by percentage, that the challenges related to time management have increased quite a lot when returning from online (high proportion of time management challenges in an onsite environment - 26.83\% -  after transitioning from online - 12.9\%). We thus noted a more difficult adaptation of students to the onsite way of working after returning from online.

\textbf{The lack of engagement is masked by organizational issues generated by the transition between onsite and online.}
It came as a surprise that there were fewer difficulties associated with a lack of engagement, drive, or commitment whether moving from onsite to online or from online to onsite. This was true in both directions (the percentage dropped from 8.93\% to 3.19\% in the first case, and it dropped from 8.79\% to 4.88\% in the second case).
These issues are easier to spot in a team which activates in a more stable and consistent environment.
Our research led us to the conclusion that when a team is dealing with organizational problems that were caused by external forces, a lesser level of commitment can be more easily concealed by these obstacles. 
\newline
\newline
\textbf{$RQ_{2}$}.\textit{Which are the effects of the online learning experience on the perception of face-to-face learning?} 

In all editions of this course, the specific learning activities could be summarized as project development, team organization, mentoring, and workshops. Students were required to develop a project (an application) working in a team, guided by a mentor from the industry and benefiting from additional training in the form of workshops. As a consequence, we will address the effects of the online learning experience on these activities.

The part of the survey asking \textit{''...aspects that made it easier to organize the team and manage the project online''} in the 2021-2022 iteration of the course led us to the following observations:
\textbf{Project development aspects:} By far the most important action, mentioned by almost all teams, was the use of tools and online platforms for different purposes:
\begin{itemize}
\item project management (in 21 cases), facilitated by tools like Jira and Trello: \textit{''organizing and sharing workload'', '' list our tasks and tick off the completed ones, so everyone could see the progress of the project'', '' allocate tasks between us''};
\item versioning (in 18 cases): GitHub was unanimously chosen as versioning system, assisting \textit{''to understand what changes were made at a certain point in the project'', ''help the team see changes in as real time as possible'', ''have access to all source code''};
\item communication - for team organization.
\end{itemize}

Another aspect that was mentioned is collaboration (in 8 cases) in project development by sharing resources and project progress, respectively to solve issues: \textit{''Mechanisms such as screen sharing have facilitated peer programming, making collaboration and code debugging much easier'', ''share screen and also IntelliJ Idea allows us to use a feature called Code allowed us to work together, live on the same system'', ''share screen, it was easy for us to see the source of the problem or look for solutions at the same time'', ''shared access to all documents''}.

\textbf{Team organization aspects.} There were three main perspectives for which students found benefits in the online environment:
\begin{itemize}
\item easier organization of meetings (mentioned 18 times) due to higher availability of team members and mentors and no need for a physical space (\textit{'' we didn't need to physically move to a place to meet'', ''because we were always on hand with everything we had to do'', ''we didn't need to find a place where everyone could get to, as these meetings were just a click away''});
\item time management, as 9 teams mentioned that they saved time or effort and 4 teams considered a flexible schedule (\textit{''organise meetings in a faster way, especially for smaller groups of  only 2-3 people'', '' we didn't have to wait until we were face to face to find a solution'', ''we didn't have to travel physically to organize team meetings'',''Everyone could work whenever they wanted'', ''each teammate could do his assignments when they wanted, not bound by a specific schedule''});
\item efficient communication (18 cases) due to the use of different communication platforms such as Slack, Microsoft Teams or WhatsApp, enabling fast responses, availability of conversation history and visibility of communication threads to all team members. Students expressed opinions such as: \textit{'' it was much easier to talk to each other when we had a problem, we didn't have to wait until we were face to face to find a solution'', ''almost daily discussions in chat applications''}.
\end{itemize}

\begin{figure*}[h]
    	\centerline{
    	\includegraphics[width=0.7\textwidth]{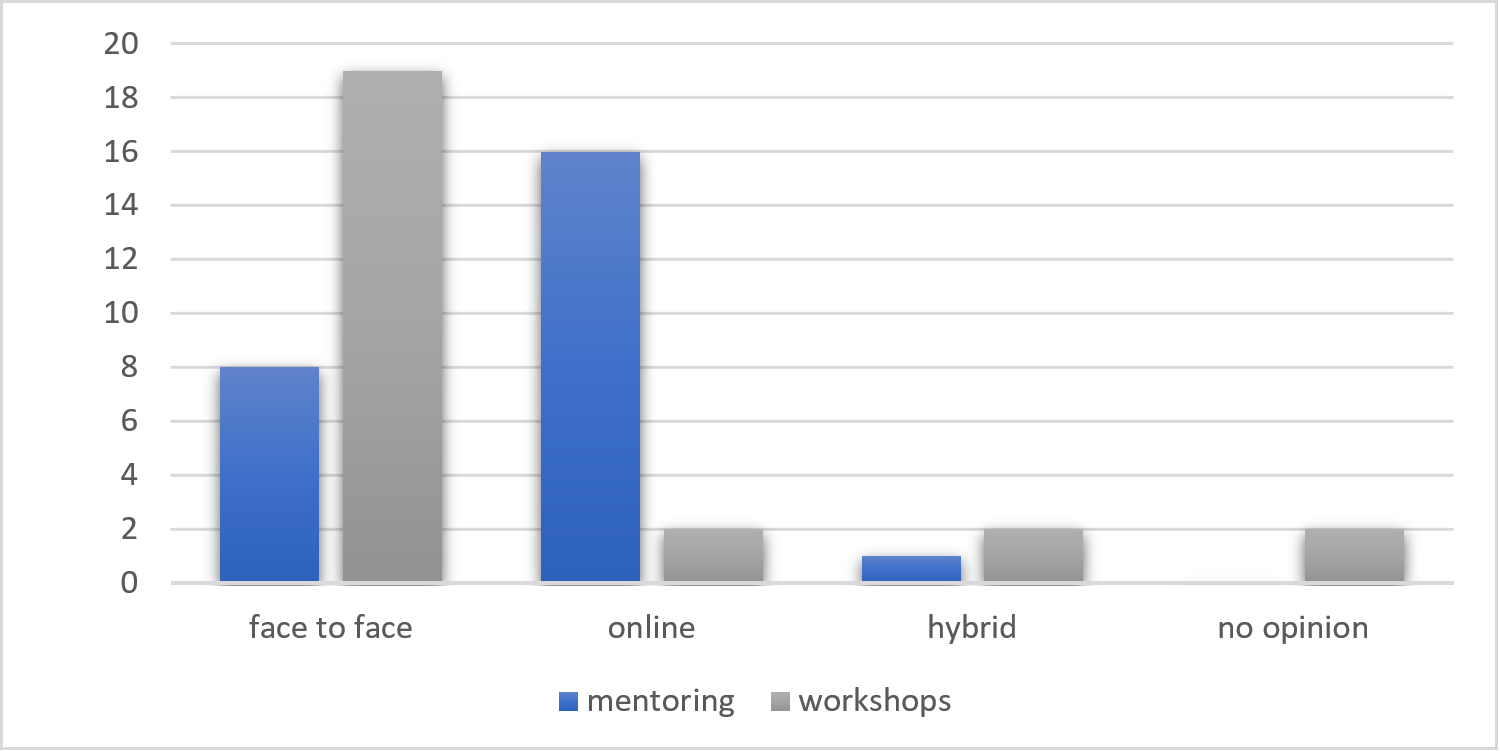}  
    }
    \caption{Students' opinions regarding online vs onsite activities}
    \label{fig:online}
\end{figure*}

In the 2022-2023 edition of the course, which comprised of complete face to face activities, we asked the students \textit{''Do you consider that participating online in meetings with mentors or workshops would have been more appropriate or, on the contrary, would have involved more disadvantages?''}. Responses have guided us to the following findings: 

\textbf{Workshops related aspects.} From 25 responses, a majority of 19 (76\%) acknowledged face to face workshops as being effective, useful, more interactive, having better engagement, less distraction and fostering networking as reflected by statements such as \textit{'' The workshops were interesting and educational, and we don't think they would have engaged us as much if they had been online'', ''The workshops worked well in-person, as it's way harder to get the audience engaged online'', ''accumulate information more efficiently''.} Two teams preferred the online environment and two teams preferred hybrid mode, arguing time spent to travel to workshop locations or scheduling restrictions, while two teams had no opinion on the subject.

\textbf{Mentoring sessions aspects.} These meetings between the industry expert and the team have the purpose of monitoring project progress and assess the students' work but also to clarify technical problems encountered by the teams. 32\% of the teams were in favor of face to face meetings, while 64\% preferred online meetings, with one team suggesting a hybrid setting. Arguments in favor of face to face activities were stated as: \textit{''face-to-face meetings can be very useful for task sharing as each member is focused on what is being discussed in that environment'', ''in physical mode was to establish new connections with specialists in the field. In addition, the face-to-face interaction strengthened relationships between team members''}. Online activities were motivated as: \textit{''better if they were online, where things such as sharing code and resources would be easier and faster, with no significant disadvantages'', ''are well suited to be held online, especially when it comes to time, as all the necessary details can easily be worked out''}

\section{Action Points}
\label{SAction}

Due to the outbreak of the Covid-19 pandemic in early 2020, all course activities within the 2020-2021 academic year were forced to take place in online environments, requiring academic staff to find appropriate solutions for effective teaching and demanding students to adapt. The challenges identified by students for this academic year were analyzed also regarding to how the transition to online unfolded. There were only a few cases (5 from a total of 35 teams) that referred to such challenges, although it should be mentioned that students already had the experience of the previous academic semester being fully online. These issues referred to the lack of face-to-face communication, lack of engagement, and teamwork in an online environment. 

The solutions found by the teaching staff together with the adaptation of students can form the basis of improving the overall learning experience even after returning to classical face-to-face academic activities. Given the results and discussions presented in the previous section, from our point of view and our expertise as academic staff, several online components can be integrated and adapted into the face-to-face scenario for team project courses:

\textbf{Development of specific materials about tools and practices for project organization:} we suggest revising the content of the workshops so that they cater specifically to the needs of students during the development of their projects. Although communication or presentation skills are very valuable in general, they do not appear as one of the most significant challenges that students confronted during the project development cycle. A topic emphasizing improved team organization and time management, including case studies, real-world examples, and industry best practices that are particular to remote project teams, is vital. This will assist students to comprehend the relevance and significance of time management and project organization skills in their area and how they may be implemented in both online and onsite environments.  

\textbf{Improve strategies for project development:} the abilities used during the pandemic for organizing and collaborating in the development of the application, without dropping the final quality of the project \cite{mot22}, strongly suggest including the use of specific tools or platforms for project management, versioning and communication as a specific requirement of the course. Even more, experience with these tools will add to the practical skills students will have on graduation, as they are the typical instruments used in real projects in software companies.

\textbf{Emphasize the importance of teamwork:} some of the main challenges students encountered in all editions of the course refer to teamwork, respectively collaboration, and leadership, which are also related to teamwork. The workshops organized as part of the course contribute to developing some skills in connection with teamwork, namely communication and an Agile mindset, but more actions can be introduced. During the first phases of the project, the mentor and workshops could concentrate on the role of the leader and try to suggest teams to adopt this position. The team leader can assume tasks related to communication, use dedicated platforms for this, and can also manage shared access to resources and documents.

As the results suggested, it might be a good decision to organize part of the meetings online. This will add to the teamwork experience and also represents an adaptation to future jobs, as many software companies encourage remote working.

\textbf{Active mentor participation:} one of the key factors for the success of this course is having industry experts as mentors. They bring their business and technical expertise and their opinion is regarded as important by students. But their time is limited and they might not have the patience or expertise typical of academic staff (for example, to assess student work in a team). In consequence, our experience during over four years of this course suggest that mentors should be selected carefully and they should assume the responsibilities of mentoring. Using online communication channels might be a good solution for increasing the mentors' availability and for providing fast feedback.

\section{Threats to validity}
\label{SThreats}
We consulted and adopted known best-practices \citep{RUN12} when designing and carrying out our investigation. We ensured that grading remained fully independent of student feedback. The report's first author acted as course coordinator across all course iterations discussed in our report. All authors have project management and software engineering experience, in addition to being experienced course coordinators in software engineering at undergraduate and master levels. The study comprised setting up the questionnaire and its collection methodology \citep{springer2022}, setting the objective and research questions, processing the free text responses through open coding and thematic analysis, as well as analyzing and interpreting the results. 

\textbf{Internal threats} were addressed by having experienced practitioners and teachers design the study and carry out the analysis. Questionnaires were validated before being handed out. Open coding and thematic analysis were carried out in pairs, with the remaining author acting to ensure consensus and validate the findings. While our study employs results that were previously analyzed \citep{enase21,springer2022,mot22}, we revisited the raw data to ensure that student feedback across all course iterations were analyzed uniformly.

With regards to \textit{construct threats}, having several course iterations and many feedback items runs the risk of having a sub-optimal number of categories, or topics per category. We managed this risk by revising all previously collected data, as well as by creating an open-data package \citep{jss2023} that other researchers can employ in order to replicate our findings or extend our study.


\textbf{External threats} relate to transferring our findings to different contexts; courses having a different curriculum, requirements or team sizes might result in different challenges and adaptations between online and in-person teaching. We manage this threat by first reporting the relevant results from the literature, and where possible, comparing the findings with those in our study. We also included detailed information regarding the target course, together with its syllabus, as well as providing the raw feedback data as collected from students. Our analysis of the literature has confirmed the existence of issues common to collaborative project-based coursework, as well as common challenges regarding the transition between in-person and online teaching \citep{cor22,springer2022}.  

\section{Conclusions and Future Work}
\label{SConclusions}

The goal of the paper was to investigate the challenges faced by students in a project based course and to evaluate the potential of online activities as instruments to be adapted in onsite courses. We analyzed student feedback given in forms of free-text survey answers during four consecutive years, including both online and face to face format. Our contributions include methodological aspects and educational insights. We have discussed the challenges in transition and the effects of online learning experience on face to face setting. Based on the results and discussion and on our experience with several iterations of this course we formulated some action points usable in project based courses.

We observed that time management and project organization were the most affected areas by the transition between on-site to online delivery, due to the difficulty of synchronizing activities and finding a common working period.

The benefits of our approach lie in reporting the impact of changes when switching from face to face to online and back to face to face education. It also serves as a guarantee for the industry that universities are considering and adapting their educational content to the future of work. With regards to open problems that remain to be addressed as well as solutions that need to be imagined, we provided our data set as a free to use replication package that we hope will encourage future investigation of the subject.

Our future plans include extending our investigation to cover other key performance indicators associated with the course, such as development process and methods that students acquire, respectively expectations versus results of their work in the frame of consecutive editions of the course. Another direction of analysis will focus on mentor participation in these courses.

\section*{Acknowledgements}

The authors express their gratitude to all companies participating in this course and contributing with valuable information. 

\scriptsize
\section*{\footnotesize Ethical issues}

Since our data collection methods involved human participants, we paid special attention to satisfy existing practices in the domain (General Data Protection Regulation (GDPR)). For the questionnaires, our procedures included a preamble to acknowledge the purpose and limitation of the content and anonymization of collected data (both personal and company information).


\printcredits

\bibliographystyle{cas-model2-names}

\bibliography{cas-refs.bib}

\begin{thebibliography}{36}
\expandafter\ifx\csname natexlab\endcsname\relax\def\natexlab#1{#1}\fi
\providecommand{\url}[1]{\texttt{#1}}
\providecommand{\href}[2]{#2}
\providecommand{\path}[1]{#1}
\providecommand{\DOIprefix}{doi:}
\providecommand{\ArXivprefix}{arXiv:}
\providecommand{\URLprefix}{URL: }
\providecommand{\Pubmedprefix}{pmid:}
\providecommand{\doi}[1]{\href{http://dx.doi.org/#1}{\path{#1}}}
\providecommand{\Pubmed}[1]{\href{pmid:#1}{\path{#1}}}
\providecommand{\bibinfo}[2]{#2}
\ifx\xfnm\relax \def\xfnm[#1]{\unskip,\space#1}\fi
\bibitem[{Adil. et~al.(2022)Adil., Fronza. and Pahl.}]{csedu22}
\bibinfo{author}{Adil., M.}, \bibinfo{author}{Fronza., I.},
  \bibinfo{author}{Pahl., C.}, \bibinfo{year}{2022}.
\newblock \bibinfo{title}{Software design and modeling practices in an online
  software engineering course: The learners’ perspective}, in:
  \bibinfo{booktitle}{Proceedings of the 14th International Conference on
  Computer Supported Education - Volume 2: CSEDU,},
  \bibinfo{organization}{INSTICC}. \bibinfo{publisher}{SciTePress}. pp.
  \bibinfo{pages}{667--674}.
\newblock \DOIprefix\doi{10.5220/0010978000003182}.
\bibitem[{{Ahmed} et~al.(2013){Ahmed}, {Capretz}, {Bouktif} and
  {Campbell}}]{FAH13}
\bibinfo{author}{{Ahmed}, F.}, \bibinfo{author}{{Capretz}, L.},
  \bibinfo{author}{{Bouktif}, S.}, \bibinfo{author}{{Campbell}, P.},
  \bibinfo{year}{2013}.
\newblock \bibinfo{title}{{Soft Skills and Software Development: A Reflection
  from Software Industry}}.
\newblock \bibinfo{journal}{Journal of Information Processing and Management}
  \bibinfo{volume}{4}, \bibinfo{pages}{171--191}.
\newblock \DOIprefix\doi{10.4156/ijipm.vol4.issue3.17}.
\bibitem[{Barr et~al.(2020)Barr, Nabir and Somerville}]{Barr20}
\bibinfo{author}{Barr, M.}, \bibinfo{author}{Nabir, S.W.},
  \bibinfo{author}{Somerville, D.}, \bibinfo{year}{2020}.
\newblock \bibinfo{title}{Online delivery of intensive software engineering
  education during the covid-19 pandemic}, in: \bibinfo{booktitle}{2020 IEEE
  32nd Conference on Software Engineering Education and Training (CSEE\&T)},
  pp. \bibinfo{pages}{1--6}.
\newblock \DOIprefix\doi{10.1109/CSEET49119.2020.9206196}.
\bibitem[{Bastarrica et~al.(2017)Bastarrica, Perovich and Samary}]{BRS17}
\bibinfo{author}{Bastarrica, M.}, \bibinfo{author}{Perovich, D.},
  \bibinfo{author}{Samary, M.}, \bibinfo{year}{2017}.
\newblock \bibinfo{title}{What can students get from a software engineering
  capstone course?}, in: \bibinfo{booktitle}{IEEE/ACM 39th ICSE-SEET}, pp.
  \bibinfo{pages}{137--145}.
\bibitem[{Braun et~al.(2019)Braun, Clarke, Hayfield and Terry}]{Braun2019}
\bibinfo{author}{Braun, V.}, \bibinfo{author}{Clarke, V.},
  \bibinfo{author}{Hayfield, N.}, \bibinfo{author}{Terry, G.},
  \bibinfo{year}{2019}.
\newblock \bibinfo{title}{Thematic Analysis}. \bibinfo{publisher}{Springer
  Singapore}.
\newblock pp. \bibinfo{pages}{843--860}.
\newblock \DOIprefix\doi{10.1007/978-981-10-5251-4\_103}.
\bibitem[{Corral and Fronza(2022)}]{cor22}
\bibinfo{author}{Corral, L.}, \bibinfo{author}{Fronza, I.},
  \bibinfo{year}{2022}.
\newblock \bibinfo{title}{It’s great to be back: An experience report
  comparing course satisfaction surveys before, during and after pandemic}, in:
  \bibinfo{booktitle}{Proceedings of the 23rd Annual Conference on Information
  Technology Education}, \bibinfo{publisher}{Association for Computing
  Machinery}, \bibinfo{address}{New York, NY, USA}. p.
  \bibinfo{pages}{66–72}.
\newblock \URLprefix \url{https://doi.org/10.1145/3537674.3554755},
  \DOIprefix\doi{10.1145/3537674.3554755}.
\bibitem[{CoT(2022)}]{cot2022}
\bibinfo{author}{CoT}, \bibinfo{year}{2022}.
\newblock \URLprefix \url{https://codeoftalent.com/}.
\bibitem[{Cruzes and Dyba(2011)}]{Cruzes}
\bibinfo{author}{Cruzes, D.S.}, \bibinfo{author}{Dyba, T.},
  \bibinfo{year}{2011}.
\newblock \bibinfo{title}{Recommended steps for thematic synthesis in software
  engineering}, in: \bibinfo{booktitle}{2011 International Symposium on
  Empirical Software Engineering and Measurement}, pp.
  \bibinfo{pages}{275--284}.
\newblock \DOIprefix\doi{10.1109/ESEM.2011.36}.
\bibitem[{Fitoussi and Chassidim(2021)}]{Fitoussi21}
\bibinfo{author}{Fitoussi, R.}, \bibinfo{author}{Chassidim, H.},
  \bibinfo{year}{2021}.
\newblock \bibinfo{title}{Teaching software engineering during covid-19
  constraint or opportunity?}, in: \bibinfo{booktitle}{2021 IEEE Global
  Engineering Education Conference (EDUCON)}, pp. \bibinfo{pages}{1727--1731}.
\newblock \DOIprefix\doi{10.1109/EDUCON46332.2021.9453896}.
\bibitem[{Gary et~al.(2017)Gary, Sohoni and Lindquist}]{Gary17}
\bibinfo{author}{Gary, K.}, \bibinfo{author}{Sohoni, S.},
  \bibinfo{author}{Lindquist, T.}, \bibinfo{year}{2017}.
\newblock \bibinfo{title}{It's not what you think: Lessons learned developing
  an online software engineering program}, in: \bibinfo{booktitle}{Proc. of
  2017 IEEE 30th Conference on Software Engineering Education and Training
  (CSEE\&T)}, pp. \bibinfo{pages}{236--240}.
\newblock \DOIprefix\doi{10.1109/CSEET.2017.45.}
\bibitem[{Gregory et~al.(2015)Gregory, Barroca, Taylor, Salah and
  Sharp}]{Peggy}
\bibinfo{author}{Gregory, P.}, \bibinfo{author}{Barroca, L.},
  \bibinfo{author}{Taylor, K.}, \bibinfo{author}{Salah, D.},
  \bibinfo{author}{Sharp, H.}, \bibinfo{year}{2015}.
\newblock \bibinfo{title}{Agile challenges in practice: A thematic analysis},
  in: \bibinfo{editor}{Lassenius, C.}, \bibinfo{editor}{Dings{\o}yr, T.},
  \bibinfo{editor}{Paasivaara, M.} (Eds.), \bibinfo{booktitle}{Agile Processes
  in Software Engineering and Extreme Programming},
  \bibinfo{publisher}{Springer}. pp. \bibinfo{pages}{64--80}.
\bibitem[{Hassan et~al.(2021)Hassan, Algahtani, Zrieq, Aldhmadi, Atta, Obeidat
  and Kadri}]{Hassan21}
\bibinfo{author}{Hassan, S.u.N.}, \bibinfo{author}{Algahtani, F.D.},
  \bibinfo{author}{Zrieq, R.}, \bibinfo{author}{Aldhmadi, B.K.},
  \bibinfo{author}{Atta, A.}, \bibinfo{author}{Obeidat, R.M.},
  \bibinfo{author}{Kadri, A.}, \bibinfo{year}{2021}.
\newblock \bibinfo{title}{Academic self-perception and course satisfaction
  among university students taking virtual classes during the covid-19 pandemic
  in the kingdom of saudi-arabia (ksa)}.
\newblock \bibinfo{journal}{Education Sciences} \bibinfo{volume}{11}.
\newblock \URLprefix \url{https://www.mdpi.com/2227-7102/11/3/134}.
\bibitem[{Hazeyama et~al.(2022)Hazeyama, Furukawa and Yamada}]{Hazeyama2022}
\bibinfo{author}{Hazeyama, A.}, \bibinfo{author}{Furukawa, K.},
  \bibinfo{author}{Yamada, Y.}, \bibinfo{year}{2022}.
\newblock \bibinfo{title}{Fully Online Project-Based Learning of Software
  Development During the COVID-19 Pandemic}. \bibinfo{publisher}{Springer
  Nature Singapore}.
\newblock pp. \bibinfo{pages}{223--232}.
\newblock \DOIprefix\doi{10.1007/978-981-16-9101-0_16}.
\bibitem[{Heberle et~al.(2018)Heberle, Neumann, Stengel and
  Regier}]{Heberle2018}
\bibinfo{author}{Heberle, A.}, \bibinfo{author}{Neumann, R.},
  \bibinfo{author}{Stengel, I.}, \bibinfo{author}{Regier, S.},
  \bibinfo{year}{2018}.
\newblock \bibinfo{title}{Teaching agile principles and software engineering
  concepts through real-life projects}, in: \bibinfo{booktitle}{2018 IEEE
  Global Engineering Education Conference (EDUCON)}, pp.
  \bibinfo{pages}{1723--1728}.
\newblock \DOIprefix\doi{10.1109/EDUCON.2018.8363442}.
\bibitem[{Holmes et~al.(2018)Holmes, Allen and Craig}]{HAC18}
\bibinfo{author}{Holmes, R.}, \bibinfo{author}{Allen, M.},
  \bibinfo{author}{Craig, M.}, \bibinfo{year}{2018}.
\newblock \bibinfo{title}{Dimensions of experientialism for software
  engineering education}, in: \bibinfo{booktitle}{Proceedings of the 40th
  ICSE-SEET}, p. \bibinfo{pages}{31–39}.
\newblock \DOIprefix\doi{10.1145/3183377.3183380}.
\bibitem[{Iacob and Faily(2019)}]{IF19}
\bibinfo{author}{Iacob, C.}, \bibinfo{author}{Faily, S.}, \bibinfo{year}{2019}.
\newblock \bibinfo{title}{Exploring the gap between the student expectations
  and the reality of teamwork in undergraduate software engineering group
  projects}.
\newblock \bibinfo{journal}{Journal of Systems and Software}
  \bibinfo{volume}{157}.
\newblock \DOIprefix\doi{https://doi.org/10.1016/j.jss.2019.110393}.
\bibitem[{Kanij and Grundy(2020)}]{Kanij20}
\bibinfo{author}{Kanij, T.}, \bibinfo{author}{Grundy, J.},
  \bibinfo{year}{2020}.
\newblock \bibinfo{title}{Adapting teaching of a software engineering service
  course due to covid-19}, in: \bibinfo{booktitle}{"2020 IEEE 32nd Conference
  on Software Engineering Education and Training (CSEE\&T)"}, pp.
  \bibinfo{pages}{1--6}.
\newblock \DOIprefix\doi{10.1109/CSEET49119.2020.9206204}.
\bibitem[{Kiger and Varpio(2020)}]{Kiger}
\bibinfo{author}{Kiger, M.E.}, \bibinfo{author}{Varpio, L.},
  \bibinfo{year}{2020}.
\newblock \bibinfo{title}{Thematic analysis of qualitative data: Amee guide no.
  131}.
\newblock \bibinfo{journal}{Medical Teacher} \bibinfo{volume}{42},
  \bibinfo{pages}{846--854}.
\newblock \DOIprefix\doi{10.1080/0142159X.2020.1755030}.
\bibitem[{Masood et~al.(2018)Masood, Hoda and Blincoe}]{Masood2018}
\bibinfo{author}{Masood, Z.}, \bibinfo{author}{Hoda, R.},
  \bibinfo{author}{Blincoe, K.}, \bibinfo{year}{2018}.
\newblock \bibinfo{title}{Adapting agile practices in university contexts}.
\newblock \bibinfo{journal}{Journal of Systems and Software}
  \bibinfo{volume}{144}, \bibinfo{pages}{501--510}.
\bibitem[{Means et~al.(2020)Means, Neisler and Associates}]{Means}
\bibinfo{author}{Means, B.}, \bibinfo{author}{Neisler, J.},
  \bibinfo{author}{Associates, L.R.}, \bibinfo{year}{2020}.
\newblock \bibinfo{title}{Suddenly online: A national survey of undergraduates
  during the covid-19 pandemic}.
\bibitem[{Motogna et~al.(2020)Motogna, Marcus and Molnar}]{Moto20}
\bibinfo{author}{Motogna, S.}, \bibinfo{author}{Marcus, A.},
  \bibinfo{author}{Molnar, A.J.}, \bibinfo{year}{2020}.
\newblock \bibinfo{title}{Adapting to online teaching in software engineering
  courses}, in: \bibinfo{booktitle}{Proceedings of the 2nd ACM SIGSOFT
  International Workshop on Education through Advanced Software Engineering and
  Artificial Intelligence}, \bibinfo{publisher}{Association for Computing
  Machinery}, \bibinfo{address}{New York, NY, USA}. p. \bibinfo{pages}{1–6}.
\newblock \URLprefix \url{https://doi.org/10.1145/3412453.3423194},
  \DOIprefix\doi{10.1145/3412453.3423194}.
\bibitem[{Motogna et~al.(2021)Motogna, Suciu and Molnar}]{enase21}
\bibinfo{author}{Motogna, S.}, \bibinfo{author}{Suciu, D.M.},
  \bibinfo{author}{Molnar, A.}, \bibinfo{year}{2021}.
\newblock \bibinfo{title}{Investigating student insight in software engineering
  team projects}, in: \bibinfo{booktitle}{Proceedings of the 16th International
  Conference on Evaluation of Novel Approaches to Software Engineering -
  ENASE}, \bibinfo{organization}{INSTICC}. \bibinfo{publisher}{SciTePress}. pp.
  \bibinfo{pages}{362--371}.
\newblock \DOIprefix\doi{10.5220/0010478803620371}.
\bibitem[{Motogna et~al.(2022a)Motogna, Suciu and Molnar}]{springer2022}
\bibinfo{author}{Motogna, S.}, \bibinfo{author}{Suciu, D.M.},
  \bibinfo{author}{Molnar, A.J.}, \bibinfo{year}{2022}a.
\newblock \bibinfo{title}{Agile mindset adoption in student team projects}, in:
  \bibinfo{editor}{Ernesto~Damiani, G.S.}, \bibinfo{editor}{Maciaszek, L.}
  (Eds.), \bibinfo{booktitle}{Evaluation of Novel Approaches to Software
  Engineering -- Revised Selected Papers (to be published)},
  \bibinfo{publisher}{Springer International Publishing},
  \bibinfo{address}{Cham}.
\bibitem[{Motogna et~al.(2022b)Motogna, Suciu and Molnar}]{mot22}
\bibinfo{author}{Motogna, S.}, \bibinfo{author}{Suciu, D.M.},
  \bibinfo{author}{Molnar, A.J.}, \bibinfo{year}{2022}b.
\newblock \bibinfo{title}{Exploring student challenges in an online
  project-based course}, in: \bibinfo{booktitle}{Proceedings of the First
  International Workshop on Designing and Running Project-Based Courses in
  Software Engineering Education}, \bibinfo{publisher}{Association for
  Computing Machinery}, \bibinfo{address}{New York, NY, USA}. p.
  \bibinfo{pages}{10–14}.
\newblock \URLprefix \url{https://doi.org/10.1145/3524487.3527361},
  \DOIprefix\doi{10.1145/3524487.3527361}.
\bibitem[{Paul and Jefferson(2019)}]{FCS}
\bibinfo{author}{Paul, J.}, \bibinfo{author}{Jefferson, F.},
  \bibinfo{year}{2019}.
\newblock \bibinfo{title}{A comparative analysis of student performance in an
  online vs. face-to-face environmental science course from 2009 to 2016}.
\newblock \bibinfo{journal}{Frontiers in Computer Science} \bibinfo{volume}{1},
  \bibinfo{pages}{7}.
\newblock \DOIprefix\doi{10.3389/fcomp.2019.00007}.
\bibitem[{Presler-Marshall et~al.(2022)Presler-Marshall, Heckman and
  Stolee}]{Presler}
\bibinfo{author}{Presler-Marshall, K.}, \bibinfo{author}{Heckman, S.},
  \bibinfo{author}{Stolee, K.T.}, \bibinfo{year}{2022}.
\newblock \bibinfo{title}{What makes team[s] work? a study of team
  characteristics in software engineering projects}, in:
  \bibinfo{booktitle}{Proceedings of the 2022 ACM Conference on International
  Computing Education Research - Volume 1}, \bibinfo{publisher}{Association for
  Computing Machinery}, \bibinfo{address}{New York, NY, USA}. p.
  \bibinfo{pages}{177–188}.
\newblock \URLprefix \url{https://doi.org/10.1145/3501385.3543980},
  \DOIprefix\doi{10.1145/3501385.3543980}.
\bibitem[{Raibulet and {Arcelli Fontana}(2018)}]{rai18}
\bibinfo{author}{Raibulet, C.}, \bibinfo{author}{{Arcelli Fontana}, F.},
  \bibinfo{year}{2018}.
\newblock \bibinfo{title}{Collaborative and teamwork software development in an
  undergraduate software engineering course}.
\newblock \bibinfo{journal}{Journal of Systems and Software}
  \bibinfo{volume}{144}, \bibinfo{pages}{409--422}.
\newblock \URLprefix
  \url{https://www.sciencedirect.com/science/article/pii/S0164121218301389},
  \DOIprefix\doi{https://doi.org/10.1016/j.jss.2018.07.010}.
\bibitem[{Raibulet and Fontana(2018)}]{RAF18}
\bibinfo{author}{Raibulet, C.}, \bibinfo{author}{Fontana, F.A.},
  \bibinfo{year}{2018}.
\newblock \bibinfo{title}{Collaborative and teamwork software development in an
  undergraduate software engineering course}.
\newblock \bibinfo{journal}{J. Syst. Softw.} \bibinfo{volume}{144},
  \bibinfo{pages}{409--422}.
\newblock \DOIprefix\doi{10.1016/j.jss.2018.07.010}.
\bibitem[{Raibulet and Lago(2022)}]{rai22b}
\bibinfo{author}{Raibulet, C.}, \bibinfo{author}{Lago, P.},
  \bibinfo{year}{2022}.
\newblock \bibinfo{title}{Industrial project-based course on service oriented
  design: Experience sharing}, in: \bibinfo{booktitle}{Proceedings of the First
  International Workshop on Designing and Running Project-Based Courses in
  Software Engineering Education}, \bibinfo{publisher}{Association for
  Computing Machinery}, \bibinfo{address}{New York, NY, USA}. p.
  \bibinfo{pages}{20–24}.
\newblock \URLprefix \url{https://doi.org/10.1145/3524487.3527360},
  \DOIprefix\doi{10.1145/3524487.3527360}.
\bibitem[{{Ralph, Paul (ed.)}(2021)}]{ACM}
\bibinfo{author}{{Ralph, Paul (ed.)}}, \bibinfo{year}{2021}.
\newblock \bibinfo{title}{{ACM Sigsoft Empirical Standards for Software
  Engineering Research}, version 0.2.0}.
\newblock \URLprefix \url{https://github.com/acmsigsoft/EmpiricalStandards}.
\bibitem[{Runeson et~al.(2012)Runeson, Host, Rainer and Regnell}]{RUN12}
\bibinfo{author}{Runeson, P.}, \bibinfo{author}{Host, M.},
  \bibinfo{author}{Rainer, A.}, \bibinfo{author}{Regnell, B.},
  \bibinfo{year}{2012}.
\newblock \bibinfo{title}{Case Study Research in Software Engineering:
  Guidelines and Examples}.
\newblock \bibinfo{edition}{1st} ed., \bibinfo{publisher}{Wiley Publishing}.
\bibitem[{Salas-Pilco(2022)}]{Salas22}
\bibinfo{author}{Salas-Pilco, S.Z.}, \bibinfo{year}{2022}.
\newblock \bibinfo{title}{The impact of covid-19 on latin american stem higher
  education: A systematic review}, in: \bibinfo{booktitle}{2022 IEEE World
  Engineering Education Conference (EDUNINE)}, pp. \bibinfo{pages}{1--6}.
\newblock \DOIprefix\doi{10.1109/EDUNINE53672.2022.9782354}.
\bibitem[{Souza et~al.(2019)Souza, Moreira and Figueiredo}]{Souza}
\bibinfo{author}{Souza, M.}, \bibinfo{author}{Moreira, R.},
  \bibinfo{author}{Figueiredo, E.}, \bibinfo{year}{2019}.
\newblock \bibinfo{title}{Students perception on the use of project-based
  learning in software engineering education}, in:
  \bibinfo{booktitle}{Proceedings of the XXXIII Brazilian Symposium on Software
  Engineering}, \bibinfo{publisher}{Association for Computing Machinery},
  \bibinfo{address}{New York, NY, USA}. p. \bibinfo{pages}{537–546}.
\newblock \URLprefix \url{https://doi.org/10.1145/3350768.3352457},
  \DOIprefix\doi{10.1145/3350768.3352457}.
\bibitem[{Suciu et~al.(2022)Suciu, Motogna and Molnar}]{jss2023}
\bibinfo{author}{Suciu, D.M.}, \bibinfo{author}{Motogna, S.},
  \bibinfo{author}{Molnar, A.J.}, \bibinfo{year}{2022}.
\newblock \bibinfo{title}{{Open Data Package for Article ''Transitioning a
  Project-Based Course between Onsite and Online. An Experience Report''}}.
\newblock \URLprefix \url{https://figshare.com/s/ae67bd64b37ccb4f2782},
  \DOIprefix\doi{10.6084/m9.figshare.21897000}.
\bibitem[{W\l{}odarski et~al.(2021)W\l{}odarski, Falleri and
  Parv\'{e}ry}]{Wloda}
\bibinfo{author}{W\l{}odarski, R.}, \bibinfo{author}{Falleri, J.R.},
  \bibinfo{author}{Parv\'{e}ry, C.}, \bibinfo{year}{2021}.
\newblock \bibinfo{title}{Assessment of a hybrid software development process
  for student projects: A controlled experiment}, in:
  \bibinfo{booktitle}{Proceedings of the 43rd International Conference on
  Software Engineering: Joint Track on Software Engineering Education and
  Training}, \bibinfo{publisher}{IEEE Press}. p. \bibinfo{pages}{289–299}.
\newblock \URLprefix \url{https://doi.org/10.1109/ICSE-SEET52601.2021.00039},
  \DOIprefix\doi{10.1109/ICSE-SEET52601.2021.00039}.
\bibitem[{Yamada et~al.(2020)Yamada, Furukawa and Hazeyama}]{Yamada20}
\bibinfo{author}{Yamada, Y.}, \bibinfo{author}{Furukawa, K.},
  \bibinfo{author}{Hazeyama, A.}, \bibinfo{year}{2020}.
\newblock \bibinfo{title}{Conducting a fully online education of a software
  engineering course with a web application development component due to the
  {COVID-19} pandemic and its evaluation}, in: \bibinfo{editor}{Wadhwa, B.},
  \bibinfo{editor}{Chawla, S.}, \bibinfo{editor}{Gan, B.},
  \bibinfo{editor}{Ouh, E.L.}, \bibinfo{editor}{Muenchaisri, P.},
  \bibinfo{editor}{Tiwari, S.}, \bibinfo{editor}{Rathore, S.S.} (Eds.),
  \bibinfo{booktitle}{Joint Proceedings of {SEED} {\&} NLPaSE co-located with
  27th APSEC Conference 2020}, \bibinfo{publisher}{CEUR-WS.org}. pp.
  \bibinfo{pages}{20--28}.

\end{thebibliography}

\bio{}
\endbio

\endbio

\end{document}